\documentstyle[twocolumn,aps]{revtex}

\begin{document}

\draft
\title{ Black plane solutions in four-dimensional spacetimes}
\author{ Rong-Gen Cai\footnote{E-mail: cairg@itp.ac.cn} 
and Yuan-Zhong Zhang}
\address{CCAST (World Laboratory), P.O. Box 8730, 
Beijing 100080, China \\
and Institute of Theoretical Physics, Academia 
Sinica, P.O. Box 2735, 
Beijing 100080, China}
\maketitle

\begin{abstract}
The static, plane symmetric solutions and cylindrically symmetric
 solutions of Einstein-Maxwell equations with a negative 
cosmological
 constant are investigated. These black configurations are 
asymptotically 
 anti-de Sitter not only in the transverse directions, but also in 
the membrane or string directions. Their causal structure is 
similar
 to that of Reissner-Nordstr\"{o}m black holes, but their Hawking 
temperature goes with $M^{1/3}$, where $M$ is the ADM mass density.
 We also discuss the static plane solutions in 
Einstein-Maxwell-dilaton
 gravity with a Liouville-type dilaton potential. The presence of 
the dilaton field changes drastically the structure of solutions. 
They are asymptotically ``anti-de Sitter'' or ``de Sitter'' 
depending 
on the parameters in the theory.
\end{abstract}

\pacs{PACS numbers: 04.20.Jb, 97.60.Lf}

\section{introduction}

In general relativity it is a subject of long-standing interest to 
look for the exact solutions of Einstein field equations. Among 
these
 exact solutions, the black hole solutions take an important 
position
 because  thermodynamics, gravitational theory, and quantum theory 
are
 connected in quantum black hole physics. In addition, black holes 
might play an important role in developing  a satisfactory quantum 
theory
 of gravitation which does not exist today. With the investigation 
of the 
lower energy actions of string theories and supergravity theory, in 
recent years, we witnessed a rapid growth of interest in the family 
of black configurations ranging from various lower dimensional 
black 
holes, black strings to higher dimensional extended black p-branes, 
such as 2-dimensional dilaton black holes [1,2,3] and black hole 
solutions in Jackiw-Teitelboim theory [4]; 3-dimensional dilaton 
black holes [5,6] and black strings [7,8]; 4-dimensional charged 
dilaton black holes [9,10,11,12] and black 
strings [13] and higher dimensional black p-branes [10,14]. These 
black configurations broaden the family of black holes and manifest
 some new 
features.

In the framework of 4-dimensional Einstein theory of gravitation, 
it
 is well known that generic black hole solutions to 
Einstein-Maxwell 
equations are Kerr-Newman solutions, which are characterized by 
only
 three parameters: the mass, charge, and the angular momentum. It 
is 
often referred to as the non-hair conjecture of black holes. The 
Kerr-Newman spacetime is asymptotically flat. When a non-zero 
cosmological constant is introduced, the spacetime will become 
asymptotically de Sitter or anti-de Sitter depending on the sign of
the cosmological constant. Although the theory of general 
relativity 
in 3-dimensions retains the same formal structure as the one in 
4-dimensions and the Einstein equations still hold, the nature of 
the
theory in 3-dimensions is very different from that of 4-dimensional 
gravitation [15]. In the 3-dimensional spacetime the number of 
independent component of Riemann tensor is six, the same as that of
 the Ricci tensor, and the Weyl tensor vanishes identically. Thus 
the 
 stress-energy tensor has only a local effect on the curvature, the  
curvature at a point is non-zero if and only if the sress-energy
 tensor does not  vanish [16]. Therefore the pure gravity is 
identically
 flat in 3-dimensional theory of gravity. But, when some extended 
matters
  appear, the Einstein equations have solutions with cosmological 
event 
horizons, The existence of a positive cosmological constant does 
not 
changes this conclusion [16]. However, a negative cosmological 
cosmological constant will changes dramatically this situation. 
Recently, Banados, Teitelboim, and Zanelli (BTZ) [17] have found a
 family of black hole solutions in 3-dimensional Einstein gravity. 
The negative cosmological constant plays a central role in the 
existence 
of BTZ black holes.

In 4-dimensional spacetime, Horowitz and Strominger [14] showed 
that 
there does not exist static, cylindrically symmetric black string 
solutions with asymptotically flat in the transverse directions if 
the 
strong energy condition,
 $T_{\mu\nu}t^{\mu}t^{\nu}\ge \frac{1}{2}Tt^{\mu}t_{\mu} $ for all 
timelike vectors $t^{\mu}$, is satisfied. When the energy condition 
is relaxed to the weak energy condition, 
$T_{\mu\nu}t^{\mu}t^{\nu}\ge 0$,
 the conclusion still holds. However, their proof does not rule out 
the
  extisence of the asymptotically non-flat black strings. The 
Kaplor's 
black string solutions [13] in the dilaton theory of gravity is a 
manifest example, which is basically a direct product of a spinning 
BTZ black hole and a real line space [18]. Therefore the Kaplor's 
black 
strings are asymptotically anti-de Sitter in the transverse 
directions 
and flat in the string direction.

In a recent paper [19], Lemos constructed the cylindrical black 
hole 
solutions (black strings) in 4-dimensional Einstein gravity with a
 negative cosmological constant. Huang and Liang [19] further 
constructed
 the so-called torus-like black holes (with the topology
 $R^2\times S^1\times S^1$). The black string solutions of Lemos is 
 asymptotically anti-de Sitter not only in the transverse 
directions, but
 also in the string direction. One of the aims of this paper is,
 in Sec. II,  to extend the work of Lemos to the plane symmetric 
solutions and cylindrically symmetric solutions in Einstein-Maxwell
 equations with a negative cosmological constant. Here, the 
negative 
 cosmological constant plays a crucial role in these solution, as  
in 
 the BTZ black holes.
 In Sec. III, we further discuss the plane symmetric solutions in 
 Einstein-Maxwell-dilaton gravity with a Liouville-type potential 
of
 the dilaton field. The presence of the dilaton field will change 
drastically the structures and quantum properties of solutions. The 
existence of black configurations is independent on the sign of the 
``cosmological constant''.  Our conclusion and a brief discussion 
are
 included in Sec. IV.

\section{ Solutions of Einstein-Maxwell equations with a
 negative cosmological constant}

In this section, we discuss the static, plane symmetric solutions
 and cylindrically symmetric solution of the Einstein-Maxwell 
equations with a negative cosmological constant, respectively. Let
 us first consider the case of plane symmetry.

\subsection{Black plane solutions}

The Einstein-Maxwell equations have a well known solution 
possessing 
the plane symmetry. Its line element is given by [21]
\begin{eqnarray} 
ds^2&=&-\left (\frac{m}{z}+\frac{e^2}{z^2}
    \right )dt^2+\left (\frac{m}{z}+\frac{e^2}{z^2}\right 
)^{-1}dr^2
          \nonumber \\
&+&z^2( dx^2+dy^2),
\end{eqnarray}
where $m$ and $e$ are two integration constants, and 
$-\infty < t,x,y,z < \infty$. when $\frac{m}{z}+\frac{e^2}{z^2}<0$,
 the solution describes a spatially homogeneous spacetime. It is 
static
 as $\frac{m}{z}+\frac{e^2}{z^2}>0$. In the latter case, the 
solution 
(1) has four Killing vectors: a timelike $\partial/\partial t$ and 
three spacelike $\partial/\partial x$, $\partial/\partial y$, and 
$x \partial/\partial y-y\partial/\partial x$, indicating the static 
plane symmetry of spacetime (1). But, the solution is of a naked 
singularity at $z=0$ and its physical meanings are unclear. On the 
other hand, Einstein equations with a negative cosmological 
constant
 ($3\alpha ^2$) admit the plane symmetric anti-de Sitter solution,
\begin{equation}
ds^2=-\alpha ^2 z^2dt^2 +(\alpha ^2 z^2)^{-1}dz^2 +\alpha ^2 z^2 
(dx^2+dy^2).
\end{equation}
If one redefines $Z=-1/(\alpha ^2 z^2)$, Eq. (2) then becomes
\begin{equation}
ds^2=(\alpha Z)^{-2}(-dt^2 +dZ^2 +dx^2 +dy^2),
\end{equation}
which is just the half of the spacetime of supergravity domain
 walls [22,23], because the gravitational field of the supergravity
 domain walls can be interpreted in terms of the domain wall 
interpolating between the Minkowski spacetime and the  plane
anti-de Sitter spacetime. The spacetime (3) is geodesically 
incomplete
 and has a Cauchy horizon at $Z\rightarrow -\infty$. This Cauchy 
horizon 
is also unstable [24], as the Cauchy horizon in the
 Reissner-Nordstr\"{o}m black holes.

Combining Eqs. (1) and (2), we find that a static plane symmetric
 solution possessing event horizons will appear. The singularity at
 the $z=0$ plane in (1) will be enclosed by these event horizons. 
We
 start with the following action
\begin{eqnarray}
S&=&\frac{1}{16\pi}\int _{V}d^4x\sqrt{-g}(R+6\alpha ^2 
-F^{\mu\nu}F_{\mu\nu}) \nonumber\\
&-&\frac{1}{8\pi}\int _{\partial V}d^3x\sqrt{-h}K,
\end{eqnarray}
where $R$ is the scalar curvature, $F_{\mu\nu}$ is Maxwell field, 
and
 $3\alpha ^2 =-\Lambda>0$ denotes the negative cosmological 
constant.
 The quantity $h$ is the induced metric on $\partial V$, and $K$ 
its
 extrinsic curvature. Varying the action (4) yields the equations 
of
 motion,
\begin{eqnarray}
G_{\mu\nu}&\equiv&R_{\mu\nu}-\frac{1}{2}Rg_{\mu\nu}=3\alpha ^2
g_{\mu\nu}
+8\pi T^{EM}_{\mu\nu},\\
0&=&\partial _{\mu}(\sqrt{-g}F^{\mu\nu}), \ \ \ 
F_{\mu\nu,\rho}+F_{\nu\rho,\mu}+F_{\rho\mu,\nu}=0,
\end{eqnarray}
where
\begin{equation}
T^{EM}_{\mu\nu}=\frac{1}{4\pi}\left 
(F_{\mu\lambda}F_{\nu}^{\lambda}-\frac{1}{4}g_{\mu\nu}F^2\right ),
\end{equation}
is the stress-energy tensor of the Maxwell field. The general 
metric
 of static plane symmetry can be written as
\begin{equation}
ds^2=-A(r)dt^2+B(r)dr^2+C(r)(dx^2+dy^2),
\end{equation}
where we have taken $r=|z| $ because of the reflection symmetry 
with 
respect  to the $z=0$ plane. In the metric (8) solving Eqs. (5) and 
(6), we find
\begin{eqnarray}
A(r)&=& B^{-1}(r)= \alpha ^2 r^2- \frac{m}{r}+\frac{q^2}{r^2},\\
C(r)&=& \alpha ^2 r^2,\\
F_{tr}&=& \frac{q}{\alpha ^2 r^2},
\end{eqnarray}
where $m$ and $q$ are two integration constants related to the ADM 
mass 
and electric charge of the solutions, respectively. Because of the 
noncompactibility of the coordinates $x$ and $y$, for simplicity, 
we 
only consider the mass and charge per unit area in the x-y plane. 
The 
electric charge density $Q$ can be obtained by the Gauss theorem,
\begin{equation}
Q=\frac{1}{4\pi}\int F_{tr} C(r)dxdy=\frac{q}{2\pi},
\end{equation}
where we have considered the two integral surfaces at $z=\pm r$. 
With
 the help of the Euclidean action method of black membranes [25], 
the
 ADM mass density $M$ is found to be $M=\alpha ^2m/4\pi$. Thus, we 
obtain the static plane symmetric solutions of Eqs. (5) and (6),
\begin{eqnarray}
ds^2&=&-\left (\alpha ^2 r^2-\frac{4\pi M}{\alpha ^2 r}+
\frac{(2\pi Q)^2}{\alpha ^4 r^2}\right )dt^2\nonumber\\
&+&\left (\alpha ^2 r^2
-\frac{4\pi M}{\alpha ^2 r}+\frac{(2\pi Q)^2}{\alpha ^4 r^2}
\right )^{-1}dr^2 + \alpha^2 r^2(dx^2+dy^2).
\end{eqnarray}
This solution is asymptotically anti-de Sitter not only in the
 transverse directions, but also in the membrane directions. By 
calculating the scalar curvature invariants in   the spacetime 
(13), 
the solution (13) is singular only at $r=0$ plane. The vacuum 
background ($M=Q=0$) corresponding to Eq. (13) is
\begin{equation}
ds^2=-\alpha ^2r^2dt^2+(\alpha ^2r^2)^{-1}dr^2+
\alpha ^2r^2(dx^2+dy^2),
\end{equation}
which is just the plane anti-de Sitter spacetime (2).
In addition, if the condition
\begin{equation}
Q^2\le\frac{3\alpha ^6}{4\pi ^2}\left (\frac{\pi M}{\alpha ^4}
\right )^{4/3}
\end{equation}
is satisfied, the equation $A(r)=0$, i.e.,
\begin{equation}
\alpha ^2 r^2-\frac{4\pi M}{\alpha ^2 r}+\frac{(2\pi Q)^2}
{\alpha ^4 r^2}=0
\end{equation}
has two positive real roots,
\begin{equation}
r_{\pm}=\frac{1}{2}\left (\sqrt{2R}\pm \left (-2R+\frac{8\pi M}
{\alpha ^4\sqrt{2R} }\right )^{1/2}\right ),
\end{equation}
where
\begin{eqnarray}
R&=&\left (\frac{\pi ^2 M^2}{\alpha ^8}+\left (\left (\frac{\pi ^2 
M^2}
{\alpha ^8}\right    )^2-\left ( \frac{4\pi ^2 Q^2}
{3\alpha ^6}\right )^3
\right )^{1/2}\right )^{1/3} \nonumber\\
 &+&\left (\frac{\pi ^2 M^2}{\alpha ^8}-\left (\left 
(\frac{\pi ^2 M^2}{\alpha ^8}\right    )^2-\left ( \frac{4\pi ^2
 Q^2}{3\alpha ^6}\right )^3\right )^{1/2}\right )^{1/3}.
\end{eqnarray}
The other two roots of Eq. (16) are image and have no physical 
meanings.
 Because $A(r)\ge 0$ when $0 \le r \le r_-$ and
 $r \ge r_+ $, $ A(r)\le 0$ when $r_-\le r \le r_+$. Therefore, the 
two
 positive roots can be interpreted as the outer horizon and inner 
horizon of the plane symmetric solutions, respectively. The 
causal structure of solution (13) is similar to that of
 Reissner-Nordstr\"{o}m black holes. The singularity at $r=0$ is
 enclosed by event horizons. Unlike the spherically symmetric black
 holes, here the singularity is in the plane at $r=0$. In addition,
 it is worth noting that, in fact, the solution (13) has four event
 horizons, two outer horizons at $z=\pm r_+$, two inner horizons
 $z=\pm r_-$. The singularity at the plane $z=0$ is enclosed by
 these horizons.  When the equality in Eq. (15) holds, the two 
horizons coincide and the horizon becomes
\begin{equation}
r_{\rm ext.}=\left (\frac{\pi M}{\alpha ^4}\right )^{1/3}.
\end{equation}
This case corresponds to the extremal black plane solutions. Here, 
we point out that if the negative cosmological constant is replaced
 by a positive one, the solution (13) will become asymptotically de 
Sitter, and has a single cosmological horizon. The singularity at
 $r=0$ becomes a cosmological singularity. We have no interest for
 this situation and do not discuss it in detail.

We now turn to thermodynamics of the black plane solutions. To do 
this, it  is convenient to employ the Euclidean method [25,26,27]. 
Analytically extending the solution (13) to its Euclidean section, 
we obtain 
\begin{eqnarray}
ds^2&=&\left (\alpha ^2 r^2-\frac{4\pi M}{\alpha ^2 r}+
\frac{(2\pi Q)^2}{\alpha ^4 r^2}\right )d\tau ^2 \nonumber\\
&+&\left 
(\alpha ^2 r^2-\frac{4\pi M}{\alpha ^2 r}+\frac{(2\pi Q)^2}
{\alpha ^4 r^2}\right )^{-1}dr^2 + \alpha^2 r^2(dx^2+dy^2),
\end{eqnarray}
where $\tau$ is the Euclidean time. The requirement of the absence 
of the conical 
singularity in the Euclidean spacetime (20) causes the Euclidean
 time $\tau$ must has a period $\beta _{H}$, which satisfies
\begin{equation}
\beta ^{-1}_H=\frac{A'}{4\pi\sqrt{AB}}\left |_{r_+} =\frac{1}{2\pi}
\left (\alpha ^2r_++\frac{2\pi M}{\alpha ^2 r_+^2}-\frac{(2\pi 
Q)^2}
{\alpha ^4 r^3_+}\right ) \right.,
\end{equation}
which is just the Hawking temperature of the black plane solutions. 
For
 extremal black planes (19) the temperature vanishes. When $Q=0$, 
that
 is, for neutral black plane solutions, the temperature, 
  $\beta ^{-1}_H=3M^{1/3}(\alpha/4\pi)^{2/3}$, goes with $M^{1/3}$, 
  which is
 very different from that of Schwarzschild black holes. It implies 
that the difference in topology structures will changes greatly the
 quantum properties of black configurations. Following Ref. [27], 
consider the black membrane and its surroundings contained by two 
infinite parallel plates at $z_B=\pm r_B$. We regard the interior 
of  the plates as the thermodymaical system and the two plates as 
the boundary. In a grand canonical ensemble, we must fix the 
boundary
 conditions of the system. The Euclidean manifold (20) is regular 
with a product topology $R^2\times R^2$ , and boundary $S^1\times 
R^2$.
 On the boundary the inverse temperature $\beta _B$ has the 
Toman's relation,
\begin{equation}
\beta _B=\beta _H A^{1/2}(r_B),
\end{equation}
and the electric potential $\phi _B$ is fixed as
\begin{equation}
\phi _B=\frac{2\pi Q}{\alpha ^2}\left (\frac{1}{r_+}-\frac{1}{r_B}
\right )A^{-1/2}(r_B).
\end{equation}
The Euclidean action can be obtained by Euclideanizing the action 
(4),
\begin{eqnarray}
S_E&=&-\frac{1}{16\pi}\int _{V}d^4x\sqrt{-g}(R+6\alpha ^2 
-F^{\mu\nu}F_{\mu\nu})\nonumber\\
&+&\frac{1}{8\pi}\int _{\partial V}d^3x\sqrt{-h}K.
\end{eqnarray}
In the Euclidean spacetime of Eq. (8), the scalar curvature and 
extrinsic curvature are, respectively,
\begin{equation}
R=-g^{-1/2}(g^{1/2}A'/AB)'-2G_0^{\ \ 0}
\end{equation}
and 
\begin{equation}
K=-g^{-1/2}(g^{1/2}B^{-1/2})'
\end{equation}
where $G_0^{\ \ 0}$ is the $0-0$ component of the Einstein tensor.
 Similar to the mass and charge, we only calculate the Euclidean 
action per unit area in the  following discussions. Substituting 
Eqs. (25) and (26) into Eq. (23), with the help of Eqs. (9)-(11),
 we have
\begin{eqnarray}
S_E&=&\beta _B\left (-\frac{\alpha ^2 r}{2\pi}A^{1/2}(r)\right )
\left |_{r_B} -\frac{\alpha ^2 r_+^2}{2}\right.\nonumber\\
&-&\left.\beta _B\frac{2\pi Q^2}
{\alpha ^2}\left (\frac{1}{r_+}-\frac{1}{r_B}\right )
A^{-1/2}(r_B) \right..
\end {eqnarray}
In order to obtain the Euclidean action of black plane solutions,
 we must eliminate the contribution of the vacuum background (14) 
with the same boundary conditions from the action (27).  The 
Euclidean action of the vacuum background is easy to get,
\begin{equation}
S_{VE}=-\beta _B \frac{\alpha ^3 r_B^2}{2\pi}.
\end{equation}
Thus, we obtain the Euclidean action of black plane solutions,
\begin{eqnarray}
S_{ME}&=&\frac{\beta _B\alpha ^2 r}{2\pi}\left (\alpha r-A^{1/2}(r)
    \right ) \left |_{r_B}-\frac{\alpha ^2 
r_+^2}{2}\right.\nonumber\\
&-&\left.\beta _B\frac{2\pi Q^2}{\alpha ^2}\left (\frac{1}{r_+}-
   \frac{1}{r_B}\right )A^{-1/2}(r_B)\right. .
\end{eqnarray}
Comparing the Euclidean action (29) with the formula of
 thermodynamic potential, we get the internal $E$, entropy $S$,
 and the chemical potential $\mu$ corresponding to the electric 
charge $Q$, respectively,
\begin{eqnarray}
E&=&\frac{\alpha ^2}{2\pi}(\alpha r^2-rA^{1/2}(r))| _{r_B},\\
S&=&\frac{\alpha ^2 r_+^2}{2}=\frac{\sigma}{4},\\
\mu &=&\frac{2\pi Q}{\alpha 
^2}(\frac{1}{r_+}-\frac{1}{r_B})A^{-1/2}
(r_B)\equiv \phi _B,
\end{eqnarray}
where $\sigma =2\alpha ^2 r_+^2$ denotes the area of horizon of 
black
 membranes and the prefactor 2 is because of two outer horizon 
surfaces. From Eq. (31) we see the entropy $S$ also satisfies 
$\frac{1}{4}$  area formula of black hole entropy. In terms of the
 relativistic thermodynamics, the proper energy $E^{\ast}$ and 
proper chemical potential $\mu ^{\ast}$ are , respectively,
\begin{eqnarray}
E^{\ast}&=&EA^{1/2}(r_B)
\stackrel{r_B\rightarrow \infty}{\longrightarrow}=M\nonumber\\
&=&{\rm ADM \ mass \ density\ of \ the\  black\  membranes},\\
\mu ^{\ast}&=& \mu A^{1/2}(r_B)
\stackrel{r_B\rightarrow \infty}{\longrightarrow}=\phi 
_H\nonumber\\
&=&{\rm electric \ potential \ at \ the \ horizon},
\end{eqnarray}
where $\phi _H=\frac{2\pi Q}{\alpha ^2}\frac{1}{r_+}$ .  With the
 help of Eqs. (30)-(32), we easily obtain the first law of
 thermodynamics for the system,
\begin{eqnarray}
dE&\equiv &\left (\frac{\partial E}{\partial S}\right 
)_{Q,\sigma}dS
+\left (\frac{\partial E}{\partial Q}\right )_{S,\sigma}dQ+\left
 (\frac{\partial E}{\partial \sigma }\right 
)_{S,Q}d\sigma,\nonumber\\
&=&\beta _B^{-1}dS+\phi _B dQ -Pd\sigma,
\end{eqnarray}
where $p\equiv -(\frac{\partial E}{\partial \sigma})_{S,Q}$ is the 
surface pressure of the system and $\sigma=2\alpha ^2 r_B^2$ the 
surface area of the system. If one rewrites  Eq. (35) by using 
proper
 quantities at $r_B\rightarrow \infty$, Eq. (35) then reduces to
\begin{equation}
dM=\beta _H^{-1}dS+\phi _HdQ,
\end{equation}
which is just the first law of black hole thermodynamics.

To end this subsection, we write down the metric of charged black 
plane solutions with a pressureless null radiation in the advanced 
time coordinates
\begin{eqnarray}
ds^2&=&-\left (\alpha ^2 r^2-\frac{4\pi M(v)}{\alpha ^2 r}+
\frac{(2\pi Q(v))^2}{\alpha ^4 r^2}\right ) dv^2\nonumber\\
&+&2dvdr +\alpha ^2 r^2 (dx^2+dy^2),
\end{eqnarray}
which is the Vaidya-like metric of black membranes. Eq. (37) 
implies 
that the stress-energy tensor of the radiation is
\begin{equation}
T^R_{\mu\nu}=\rho (v,r)l_{\mu}l_{\nu},
\end{equation}
where $l_{\mu}=-\partial _{\mu} v$ is the four-velocity of the null 
radiation, and the energy density $\rho (v,r)$ satisfies
\begin{equation}
\rho (v,r)=\frac{\dot{M}(v)}{4\alpha ^2 r^2}-\frac{\pi Q\dot{Q}(v)}
{\alpha ^4 r^3},
\end{equation}
where an overdot stands for derivative with respect to $v$. 
Following
 Refs. [28,29,30], we can easily show that the inner horizon $r_-$ 
is 
unstable.  When the ingoing radiation has a power-law tail, a 
nonscalar curvature singurality will be developed at the inner 
horizon. 
When an
 outgoing null flux is added to the metric (37), the mass inflation 
will take place inside the black plane solutions, as in the
 Reissner-Nordstr\"{o}m black holes.

\subsection{Black string solutions}

The black string solutions to Einstein equations with a negative 
cosmological constant have been constructed by Lemos [19],
\begin{eqnarray}
ds ^2&=&-\left (\alpha ^2 r^2-\frac{m}{r}\right )dt^2\nonumber\\
&+&
 \left (\alpha ^2 r^2-\frac{m}{r}\right )^{-1}dr^2 
+r^2 d\theta ^2 +\alpha ^2 r^2 dz^2,
\end{eqnarray}
where $-\infty <t,z<\infty$, $0\le r<\infty$, and $0\le \theta \le 
2\pi$,
 the integration constant $m$ is related to the ADM mass density of 
the
 black strings. Huang [31] has recently discussed the the 
generalization of the solutions (40) to including the electric 
charge.
 However, some expressions in Ref.[31] are incorrect. For 
completeness,
 here we reexamine the charged black string solutions to 
Einstein-Maxwell
  equations with a negative cosmological constant. To construct the 
cylindrically symmetric solution, By identifying the coordinate $x$ 
in Eq. (6) with a period $2\pi$, and replacing the variable $y$ by 
$z$, 
 we obtain the static cylindrically symmetric solution of Eqs. (5) 
and (6),
\begin{eqnarray}
ds^2&=&-\left (\alpha ^2 r^2-\frac{4M}{\alpha r}
\frac{4Q^2}
{\alpha ^2r^2}\right )dt^2\nonumber\\
&+&\left (\alpha ^2 r^2-\frac{4M}{\alpha r}+\frac{4Q^2}
{\alpha ^2r^2}\right )^{-1}dr^2 +r^2d\theta ^2 +
\alpha ^2 r^2 dz^2,\\
F_{tr}&=&\frac{2Q}{\alpha r^2},
\end{eqnarray}
where the two constants $M$ and $Q$ are the ADM mass and charge per
 unit length in the $z$ direction. The spacetime (41) is 
asymptotically
 anti-de Sitter in the transverse directions and string directions,
 unlike the Kaplor's black strings in the dilaton gravity [13]. The 
singularity at $r=0$ is enclosed by the horizons $r_{\pm}$ if the 
condition
\begin{equation}
Q^2\le \frac{3}{4}M^{4/3}
\end{equation}
holds.  As same as the black plane solutions, the black strings 
(41)
 have two horizons,
\begin{equation}
r_{\pm}=\frac{1}{2}\left (\sqrt{2R}\pm \left (-2R+\frac{8M}
{\alpha ^3\sqrt{2R}}\right )^{1/2}\right ),
\end{equation}
where
\begin{eqnarray}
R&=&\left (\frac{M^2}{\alpha ^6}+\left (\left (\frac{M^2}{\alpha 
^6}
\right )^2-\left (\frac{4Q^2}{3\alpha ^4}\right )^3\right )^{1/2}
\right )^{1/3}\nonumber\\
&+&\left (\frac{M^2}{\alpha ^6}-\left (\left (\frac{M^2}
{\alpha ^6}\right )^2-\left (\frac{4Q^2}{3\alpha ^4}\right )^3
\right )^{1/2}\right )^{1/3}.
\end{eqnarray}
Euclideanizing the metric (41), we can get the Hawking temperature 
of
 the black strings,
\begin{equation}
\beta ^{-1}_H=\frac{1}{2\pi}\left (\alpha ^2 r_++\frac{2M}{\alpha 
r_+^2}
-\frac{4Q^2}{\alpha ^2r_+^3}\right ).
\end{equation}
Similar to the previous subsection, we have the entropy per unit 
length 
and the first law of thermodynamics for charged black strings,
\begin{eqnarray}
S&=&\frac{\pi \alpha r_+^2}{2}=\frac{1}{4}\sigma,\\
dM&=&\beta ^{-1}_HdS+\phi _HdQ,
\end{eqnarray}
where $\sigma=2\pi \alpha r_+^2$ is the area of horizon per unit 
length
 and $\phi _H=\frac{2Q}{\alpha r_+}$ the electric potential at the
 horizon $r_+$. When $Q=0$, i.e., for neutral black strings, the 
inner
 horizon disappears and $r_+=\alpha ^{-1}(4M)^{1/3}$. Thus the 
Hawking 
 temperature (46) becomes $\beta ^{-1}_H=\frac{3\alpha}{2\pi} \left
  (\frac{M}{2}\right )^{1/3}$. Therefore, the temperature of black
 strings also goes with $M^{1/3}$, as the case of black membranes.  
When the equality in Eq. (43) holds, the two horizons coincide and 
Hawking temperature vanishes. This corresponds to the extremal 
black 
strings. As the case of back membranes, the causal structure of
 charged black strings is similar to that of Reissner-Nordstr\"{o}m 
black holes.

Finally, we write down here the Vaidya-like metric of black
 strings (41),
 \begin{eqnarray}
ds^2&=&-\left (\alpha ^2 r^2-\frac{4M(v)}{\alpha r}+\frac{4Q^2(v)}
{\alpha ^2r^2}\right )dv^2\nonumber\\
&+&2dvdr+r^2d\theta ^2 +\alpha ^2 r^2dz^2,
\end{eqnarray}
where the energy density of the null radiation is
\begin{equation}
\rho (v)=\frac{\dot{M}(v)}{2\pi\alpha r^2}-\frac{Q\dot{Q}(v)}{\pi
 \alpha ^2 r^3}.
\end{equation}
Similarly,  by using the metric (49) we can show that the inner
 horizon also unstable and a scalar curvature singularity will 
replace the inner horizon when
the charged black strings are perturbed by ingoing and outgoing 
null fluxes.

So far, we have investigated the static, plane symmetric solutions
 and cylindrically symmetric solutions in the Einstein-Maxwell 
equations with a negative cosmological constant.  The causal 
structure of these solutions is similar to that of 
Reissner-Nordstr\"{o}m black holes. Therefore they can be 
interpreted
 as the black membranes and black strings, respectively. These 
black
  configurations are asymptotically anti-de Sitter not only in the 
transverse directions, but also in the membrane or string 
directions. 
 In these solutions, the negative cosmological constant plays an 
important role. In the following section, we will see that when the 
dilaton field is present, the structure of the plane solutions will
 be changed greatly. The role of the negative cosmological constant
seems to be lowered.

\section{Black plane solutions in Einstein-Maxwell-dilaton gravity}

In recent years, many black hole solutions have been found in the
 dilaton gravity. Due to the dilaton field, the usual black hole 
structure and quantum 
properties are changed drastically. In this section, we would like
 to look for the plane symmetric solution in the
 Einstein-Maxwell-dilaton gravity with a Liouville-type dilaton 
potential, whose action is
\begin{equation}
S=\frac{1}{16\pi}\int d^4x\sqrt{-g}\left[ R-2(\nabla \phi)^2
-6\alpha ^2\eta
e^{2b\phi}-e^{-2a\phi}F^2\right ],
\end{equation}
where $\phi$ is the dilaton field, the Liouville-type potential 
represents the ``cosmological constant term'', $a$ and $b$ are 
two constants, and $\eta=\pm 1$, representing the sign of the 
``cosmological constant''. This action (51) has been considerably
 investigated in the context of 3- and 4-dimensional dilaton 
black holes [5,12].  Varying the action (51), we obtain the 
equations of motion,
\begin{eqnarray}
R_{\mu\nu}&=& 2\partial _{\mu}\phi \partial_{\nu}\phi+
3\alpha ^2\eta e^{2b\phi}g_{\mu\nu}\nonumber\\
&+&2e^{-2a\phi}(F_{\mu\lambda}F_{\nu}^{\lambda}-
\frac{1}{4}g_{\mu\nu}F^2 ),\\
0&=&\partial_{\mu}(\sqrt{-g}e^{-2a\phi }F^{\mu\nu}),\\
\nabla ^2\phi &=&3\alpha ^2 b\eta e^{2b\phi}-\frac{a}{2}
e^{-2a\phi}F^2.
\end{eqnarray}
We consider again the static plane solutions of Eqs. (52)-(54) 
in the following
metric
\begin{equation}
ds^2=-A(r)dt^2+B(r)dr^2 +C(r)(dx^2+dy^2).
\end{equation}
From (53) we have
\begin{equation}
F_{tr}=\frac{Q}{\sqrt{AB}C}e^{2a\phi},
\end{equation}
where $Q$ is an integration constant. Thus, Eqs. (52) and (54) 
reduce to
\begin{eqnarray}
&&\frac{A'C'}{2ABC}-\frac{C''}{BC}+\frac{C'^2}{2BC^2}
+\frac{B'C'}{2B^2C}
=\frac{2}{B}\phi '^2,\\
&&\frac{A''}{2AB}-\frac{A'B'}{4AB^2}-\frac{A'^2}{4A^2B}
+\frac{A'C'}{2ABC}=-3\alpha ^2\eta e^{2b\phi}\nonumber\\
&&+\frac{Q^2}
{A^2B^2C^2}e^{2a\phi},\\
&&-\frac{C''}{2BC}+\frac{B'C'}{4B^2C}-\frac{A'C'}{4ABC}=
3\alpha ^2\eta e^{2b\phi}+\frac{Q^2}{A^2B^2C^2}e^{2a\phi},\\
&&\frac{1}{\sqrt{AB}C}\left [\sqrt{AB}\left(\frac{C}{B}\right)
\phi '\right ]'=3\alpha ^2b\eta e^{2b\phi}+\frac{aQ^2}
{A^2B^2C^2}e^{2a\phi},
\end{eqnarray}
where a prime denotes derivative with respect to $r$.  From
 Eqs. (57)-(60), we obtain a set of solutions,
\begin{eqnarray}
A(r)&=&\frac{1}{B(r)}=-\frac{4\pi M}{N \alpha ^N}r^{1-N}
-\frac{6\alpha ^2\eta}{N(2N-1)}r^N \nonumber\\
&+& \frac{2Q^2}
{N\alpha ^{2N}}r^{-N},\\
C(r)&=&(\alpha r)^N,\\
\phi (r)&=&-\frac{\beta}{2}\ln r,
\end{eqnarray}
where $M$ is the quasilocal mass density [32], and
\begin{eqnarray}
\beta &=&\sqrt{2N-N^2},\\
a&=&b=\beta /N.
\end{eqnarray}
In  the spacetime described by (61) and (62), the scalar curvature 
is
\begin{equation}
R=\frac{\beta ^2}{2r^2}A(r)+12\alpha ^2\eta r^{N-2}.
\end{equation}
Obviously, the curvature diverges at $r=0$. Therefore the $r=0$ 
plane is a singularity plane in solutions (61). The plane symmetric 
solutions (61) manifest some interesting properties because of the
 parameter $N$. We will separately discuss the cases of
 $\eta= -1$ and $\eta =1$.

(1) $\eta =-1$. That is, the Liouville-type potential corresponds 
to a ``negative cosmological constant''. The solution (61) is of
 different  asymptotic properties as the $N$ is taken different 
values.
 From  the solution Eqs. (61)-(65), we have $0<N<2$, but $N\neq 
1/2$.
 When $N=2$, the solution can be reduced to the one of 
Einstein-Maxwell
  equations with a negative cosmological constant (13).

(i) When $ 1/2 <N<2$, the second term ($r^N$) in Eq. (61) is 
dominant
 as $r\rightarrow \infty$. In that case,  the solution (61) is 
asymptotically ``anti-de Sitter'', where the word ``anti-de 
Sitter''
 means that the solution has no cosmological horizon. 
The other horizons are given by the equation $A(r)=0$, i.e.
\begin{equation}
\frac{3\alpha ^2}{(2N-1)}r^{2N}-\frac{2\pi M}{ \alpha ^N}r
+\frac{Q^2}{\alpha ^{2N}}=0.
\end{equation}
Because of the higher order of $r$ in Eq. (67), in general, the 
solution (61) will has the multi-horizon structures. A simper case
 is $N=1$, in this case, we have
\begin{equation}
r_{\pm}=\frac{1}{3\alpha ^3}\left (\pi M \pm \sqrt{\pi ^2 M^2
-3\alpha ^2 Q^2}\right).
\end{equation}
When $M^2>\frac{3\alpha ^2 Q^2}{\pi ^2}$, the solution (61) has two
 horizons, outer horizon $r_+$ and inner horizon $r_-$; when $M^2=
\frac{3\alpha ^2 Q^2}{\pi ^2}$, the solution (61) has a single
 horizon $r_+=\pi M/(3\alpha ^3)$, this corresponds to the extremal
 plane solution;  when $M^2<\frac{3\alpha ^2 Q^2}{\pi ^2}$, the
 solution (61) will has no horizon and the singularity at $r=0$
 becomes naked. Evidently, the causal structure of this case is
 similar to that of Reissner-Nordstr\"{o}m black holes. The
 Hawking temperature is 
\begin{equation}
\beta ^{-1}_H=\frac{1}{2\pi}\left ( 3\alpha ^2-
\frac{Q^2}{\alpha ^2r_+^2}\right ).
\end{equation}
 From (69) we can see that, if $Q=0$, the temperature is a 
constant.
 This is very different from the case in the absence of the dilaton 
field (21). For a generic $N$, the Hawking temperature is
\begin{eqnarray}
\beta _H^{-1}&=&\frac{1}{2\pi}\left (-\frac{2\pi M(1-N)}{N\alpha 
^N}
r_+^{-N} +\frac{3\alpha ^2}{(2N-1)}r_+^{N-1}\right.\nonumber\\
&-&\left.\frac{Q^2}{\alpha ^{2N}}r_+^{-N-1}\right ).
\end{eqnarray}
It should be noted that  for some special $N$, the solution (61) 
will
 has no horizon. For example, when $N=3/2$, the solution has no 
horizon, and the singularity at $r=0$ is naked.

(ii) When $0<N<1/2$, the first term  ($r^{1-N}$) in Eq. (61) is 
dominant as $r\rightarrow \infty$. In that case, the solution (61)
 is asymptotically `` de Sitter'',
 where the `` de Sitter'' means that the solution (61) has the 
cosmological horizon. In general, the plane solution will be of 
the inner horizons, outer horizon, and the cosmological horizon.
 These horizons are determined by the equation
\begin{equation}
\frac{3\alpha ^2}{(1-2N)}r^{2N}+\frac{2\pi M}{\alpha ^N}r 
-\frac{Q^2}{\alpha ^{2N}}=0.
\end{equation}
In particular, we find that, in some special cases,  although there
 exists the cosmological horizon, the inner and outer horizons are 
absent, the singularity at $r=0$ is a cosmological  singularity.  
A manifest example is $N=1/4$, the solution has only the 
cosmological
 horizon,
\begin{equation}
r_{\rm coh}=\left [\frac{\alpha ^{1/4}}{4\pi M}\left (-6\alpha^2+
(36 \alpha ^4  +8\pi MQ^2\alpha ^{-3/4})^{1/2}\right )
 \right ]^{1/2}.
\end{equation}
This situation is very like the Reissner-Nordstr\"{o}m-de Sitter
 spacetime when the charge $Q$ excess a critical value. But there
 exists  an essential difference in the causes. The former is 
purely
 because of the parameter $N$; the latter is due to the relation of
 black hole hairs (mass, charge, and cosmological constant).

(2) $\eta =1$. Namely, the Liouville-type potential corresponds to
 a ``positive cosmological constant'. In that case, 
$A(r)\rightarrow
 +\infty$ as $r\rightarrow 0$, and $A(r)\rightarrow -\infty$ as
 $r\rightarrow +\infty$, therefore, the equation $A(r)=0$ 
determining
 the horizons of solutions has at least a positive root between
 $0<r<\infty$. For $1/2<N<2$ and $0<N<1/2$, the solutions (61) are
 all asymptotically ``de Sitter'', that is,  these solutions have 
the cosmological horizons. Of course, for generic parameter $N$, 
these solutions could have  the inner horizons and outer horizon,
 indicating the multi-horizon feature. These horizons are  given by 
\begin{equation}
\frac{3\alpha ^2}{(2N-1)}r^{2N}+\frac{2\pi M}{ \alpha ^N}r
-\frac{Q^2}{\alpha ^{2N}}=0.
\end{equation}
However, unlike the case $\eta =-1$, when $N=1$, $1/4$, or $3/2$, 
the solution (61) has only a cosmological horizon, which is
\begin{equation}
r_{\rm coh}=\frac{1}{3\alpha ^3}\left (-\pi M + \sqrt{\pi ^2 M^2+
3\alpha ^2 Q^2}\right),
\end{equation}
for $N=1$;
\begin{equation}
r_{\rm coh}=\left [\frac{\alpha ^{1/4}}{4\pi M}\left (6\alpha^2+
(36 \alpha ^4  +8\pi MQ^2\alpha ^{-3/4})^{1/2}\right ) \right 
]^{1/2},
\end{equation}
for $N=1/4$;
\begin{eqnarray}
r_{\rm coh}&=&\left [ \frac{Q^2}{3\alpha ^5}+\left ( \left 
(\frac{Q^2}
{3\alpha ^5}\right )^2+\left (\frac{4\pi M}{9\alpha ^{7/2}}
\right )^3\right )^{1/2}\right ]^{1/3}\nonumber\\
&+&
\left [\frac{Q^2}{3\alpha ^5}-\left ( \left (\frac{Q^2}{3\alpha ^5}
\right )^2+\left (\frac{4\pi M}{9\alpha ^{7/2}}\right )^3\right 
)^{1/2}
\right ]^{1/3},
\end{eqnarray}
for $N=3/2$.  The Hawking temperature for these cosmological 
horizons
 is
\begin{eqnarray}
 \beta _H^{-1}&=&\frac{1}{2\pi}\left | \left (-\frac{ 2\pi M(1-N)}
{N\alpha ^N}r^{-N} -\frac{3\alpha ^2}{(2N-1)}r^{N-1}\right.\right.
\nonumber\\
&-&\left.\left.
\frac{Q^2}{\alpha ^{2N}}r^{-N-1}\right )\right | _{r=r_{\rm coh}}.
\end{eqnarray}

Similarly, the cylindrically symmetric solutions in the action (51) 
can also be obtained. The causal structures of them is similar to
 those of the  plane symmetric solutions Eqs. (61)-(65). For 
simplicity,
 here we do not present them.

\section{ Conclusion and discussions}

In this work we have discussed the static, plane symmetrically 
solutions
 and cylindrically symmetric solutions in Einstein-Maxwell 
equations 
with a negative cosmological constant. The singurality at $r=0$
 can be 
enclosed by event horizons. Their causal structure is very similar 
to 
the one of Reissner-Nordstr\"{o} black holes, but the Hawking 
temperature goes with $M^{1/3}$. These black configurations are 
asymptotically anti-de Sitter not only in the transverse 
directions,
 but also in the membrane or string directions. In these solutions 
with horizons the negative cosmological constant takes a crucial 
role, 
as in the 3-dimensional BTZ black holes.  We have also investigated 
the plane symmetric solutions in Einstein-Maxwell-dilaton gravity 
with
 a Liouville-type diatonic potential. The presence of the dilaton 
field changes drastically the structure of the solutions to 
Einstein-Maxwell equations with a cosmological constant. In 
particular, 
 there exist the black plane solutions for  the `` positive 
cosmological 
 constant'' and ``negative cosmological constant''.  These 
solutions 
 are asymptotically ``anti-de Sitter'' or ``de Sitter'', depending 
on the
 parameters $N$ and $\eta$.

In the plane symmetric solutions, an interesting phenomenon is 
that,
 if
 one removes the reflection symmetry with respect to the $z=0$ 
plane, 
the black plane solution becomes that the singurality at $z=0$ 
plane 
is enclosed by event horizon in one direction and naked in  the
 another direction. For example, for neutral plane solutions,I
\begin{eqnarray}
ds^2&=&-\left (\alpha ^2 z^2 -\frac{4\pi M}{\alpha ^2 z}\right ) 
dt^2 
\nonumber\\
&+&\left (\alpha ^2 z^2 -\frac{4\pi M}{\alpha^2 z}\right ) dz^2 +
(\alpha z)^2(dx^2 +dy^2).
\end{eqnarray}
The solution has a singularity at $z=0$ plane. When $M>0$, 
obviously,
 it has a horizon at $z=(4\pi M/\alpha ^4)^{1/3}$ in the positive
 $z$ direction. But, the singurality is naked in the negative $z$ 
direction. When $M<0$, the situation is opposite. The property is
 the new feature of these black plane solutions. Of course, the 
problems of physics might have the reflection symmetry. Finally,
 we  would like to point out that the black plane solutions have
 been also discussed by Cvetic [23] in the context of 
supergravity domain walls.

\begin{flushleft}
{\bf{\large Acknowledgements}}\\
This work was supported in part by China Postdoctoral Science 
Foundation. One of authors (R. G. Cai) would like to thank 
Dr. C. G. Huang for useful discussions.  
\end{flushleft}

\end{document}